\documentclass[debug]{rmaa}

\usepackage{paralist}

\usepackage{psfrag,color}

\usepackage[latin1]{inputenc}

\title{PAHs AS TRACERS OF LOCAL AGNs-STARBURST CONNECTION} 

\author{
  Mario-A. Higuera-G,\altaffilmark{1} 
  Andr\'es F. Ramos P.,\altaffilmark{2}
 }

\altaffiltext{1}{Observatorio Astron\'omico Nacional, Facultad de Ciencias, Universidad Nacional de Colombia.}
\altaffiltext{2}{Departamento de F\'isica, Facultad de Ciencias, Universidad Nacional de Colombia.}

\shortauthor{Higuera-G Mario-A}
\shorttitle{RevMexAA Main Journal Demo Document}

\fulladdresses{
\item Mario Armando Higuera Garz\'on: Observatorio Astron\'omico Nacional, Facultad de Ciencias, Universidad Nacional de Colombia, Bogot\'a, Colombia
  (mahiguerag@unal.edu.co).
  
  \item Andr\'es Felipe Ramos Padilla: Departamento de F\'isica, Facultad de Ciencias, Universidad Nacional de Colombia, Bogot\'a, Colombia
  (aframosp@unal.edu.co).
  }

\listofauthors{W. J. Henney, A. Collaborator, \& L. Author}
\indexauthor{Higuera-G, Mario-A.}

\abstract{The main purpose of this research was to investigate how energetic processes associated with Active Galactic Nuclei are related to those due to nuclear or circumnuclear star formation activity. Photometric and spectroscopic data were used  to discriminate these processes in a sample of starburst, infrared galaxies and AGNs. Herein, we propose new diagnostic diagrams based on the 7.7 $\mu$m polycyclic aromatic hydrocarbon emission band, the  $\textrm{L}(\textrm{MIR},\textrm{FIR})$ infrared ratio and the $q$ parameter.  The diagnostic diagrams allow us to discriminate the behavior of quasars and Seyfert 1-Seyfert 2 galaxies from starbursts and LIRGs-ULIRGs objects.}

\resumen{El objetivo central de este trabajo fue investigar c\'omo procesos energ\'eticos asociados a la actividad de los n\'ucleos activos de galaxias est\'an vinculados con aquellos que son debidos a la actividad de formaci\'on estelar nuclear y circumnuclear. Datos fotom\'etricos y espectrosc\'opicos fueron usados con el fin de discriminar estos procesos en un conjunto de galaxias starburtst, infrared galaxies y AGNs. Se proponen nuevos diagramas de diagn\'ostico basados en la emisi\'on de PAH en 7.7 $\mu$m, la raz\'on entre la emisi\'on en el infrarrojo medio y lejano $\textrm{L}(\textrm{MIR},\textrm{FIR})$ y el par\'ametro $q$. Los diagramas de diagn\'ostico, permiten comparar el comportamiento de los quasares y galaxias Seyfert 1- Seyfert 2 con las galaxias starburst y LIRGs-ULIRGs.}

\addkeyword{Active Galactic Nuclei}
\addkeyword{Star formation rate}
\addkeyword{Mid-Far infrared emission}
\addkeyword{Radio emission}

\begin{document}
\maketitle

\section{INTRODUCTION}
\label{sec:intro}

Numerous works \citep{cidf01,spoo01,iman06,maio07,alex12}, have made important contributions to quantify the star formation activity in several nuclear and circumnuclear regions of Active Galactic Nuclei (AGNs), and many diagnostic diagrams have been built in order to quantify the contribution of AGN and star formation to the IR luminosity. These diagrams use some bands of continuum and atomic emissions. The IR spectrum has many wavelengths that can be used to study  the  origin and connection between AGN activity and stellar processes \citep{clav00,stor01,iman04,davi07}. The nuclear dust in AGNs is warmed up by the radiation of the central source, also by the emissions coming from the ionized gas \citep{sand96,mele08} and by the emissions associated with stellar formation, the latter dominant in  starburst and infrared Type galaxies. \citep{bran06,iman06}.\\

 \citet{kenn98} and \citet{panu03} carried out extensive research that resulted in a self-consistent set of rate estimates of star formation along the Hubble sequence. These calibrations scale linearly with luminosity in two ranges of the continuum, ultraviolet and far infrared, as well as emission lines such as H$\alpha$, H$\beta$, P$\alpha$, P$\beta$, Br$\alpha$, Br$\gamma$ and low probability of lines [O II].  These standard SFR indicators are often dominated by the AGN itself, particularly for unobscured (i.e., Type 1) sources. To mitigate this problem, \citet{iman04,schw06} and others use Polycyclic Aromatic Hydrocarbons (PAHs) to trace star formation. The PAHs, radiated through IR fluorescence following vibrational excitation by a single ultraviolet (UV) photon, provide an indirect measurement of UV radiation field strength from stars \citep{peet02, alex12}. However, high-energy photons or shocks linked with AGN may destroy or modify their molecular carriers \citep{yong07}. Recently, \citet{maio07} and \citet{alex12} introduced two other rate calibrations for star formation using 7.7$\mu$m, and 11,3 $\mu$m polycyclic aromatic hydrocarbon (PAH) luminosities, respectively.\\

 In this paper, we describe the estimation of an MIR-FIR indicator, and calculate the $q$ parameter in order to construct two new diagnostic diagrams.\\

\section{The sample and sources}

The galaxy data collected for this work are summarize in Table \ref{datasource}. As is shown in this table, the sample is heterogeneous and came from different instruments: the IR continuum from the IRAS satellite, the radio continuum from NRAO, and the PAHs bands extracted by us from the Spitzer database. The galaxies have a redshift range between 0.001878 and 0.55. The Mid-Infrarred and Far-Infrarred observations in 12, 25, 60 and 100 $\mu$m, came from a sub sample of the complete list of objects contained in the IRAS Revised Bright Galaxy Sample (RBGS), a flux- limited sample of all extragalactic objects brighter than 5.24 Jy at 60 $\mu$m, covering the entire sky surveyed by the Infrared Astronomical Satellite. These data are condensed in IRAS catalogues \citep{neug84, soif21, mosh90, huch92, rush93, sand03}. The samples are not expected to be biased by the presence of nuclear starburst \citep{iman03,iman04}. For the same subsample of IR objects, we selected corrected flux densities of 1.4 GHz, from the complete NRAO VLA Sky Survey (NVSS) \citep{cond91, cond98}. The NVSS contains the majority of galaxies in the IRAS Faint Source Catalog. Finally,  the PAH measurements were extracted from Spitzer satellite observations. These observations use the data (BCD) of the Spitzer Heritage Archive (SHA) using 5.3-14 $\mu$m (SL) and 14-40 $\mu$m  (LL) low resolution (R $\sim$100) modules with widths of $\sim$ 3.7'' and 10.5'' respectively. The data were extracted using three algorithms in the IDL environment: CUBISM, a method for reconstructing spectral cubes using the BCD default spectral images, IRSCLEAN, which creates a mask of damaged pixels from the BCD data set, and PAHFIT, an algorithm for the decomposition of the MIR spectrum, which focuses on the overall behavior of PAH emissions relative to its underlying continuum.\\

The sample consists of one subsample composed of typical IR emission galaxies: 14 (9) LIRGs  selected from \citet{iman06b} and \citet{iman06}; 24 (21) ULIRGs-LINER, 12 (8) ULIRGs-HII, 7 (4) ULIRGs-Sy1 y 7 (5) ULIRGs-Sy2, published by \citet{iman06} and \citet{iman07}; 20 (16) starburst galaxies published by \citet{bran06}, and a second subsample of AGNs composed of 41 (27)  Seyfert 1  and 53 (47)  Seyfert 2 galaxies published by \citet{iman03}, \citet{iman04}, \citet{rodr03}, \citet{rush93}, \citet{huch92} and \citet{clav00}; 108 (14) PG and 93 (16) 3CR quasars and QSOs taken from the work of \citet{schw06} and \citet{yong07}. The values in parentheses show the galaxies extracted by us in order to get the PAH features. The information was complemented with data selected from the NASA/IPAC Extragalactic Database-NED, Simbad-VizieR database and 2MASS. Although, not all the sample are statistically complete, the subsamples provide useful information about the behavior and particularities of the different Type of galaxies selected for this work, and major conclusions can be obtained. \\


\section {The MIR-FIR indicator}\label{MidFIR}

It is well known that the emission continuum in the near, mid and far-infrared spectra, represents different temperature states of molecular clouds and their possible origins (AGN or stellar activity). For this reason, we define the measure L(12,60) in order to find differences between the emissions in the MIR and FIR, following a similar criteria adopted by \citet{mura00},

\begin{equation}
\textrm{L}(12,60)=\log\Big[\frac{\textrm{L}(12\mu \textrm{m})}{\textrm{L}(60\mu \textrm{m})}\Big] .
\label{eq1}
\end{equation}

$\textrm{L}(12.60)$ values for quasars and Seyfert 1 galaxies, here grouped together under the label \textit {Type 1}, are in the range of - 0.63 to 1.05. For all Seyfert 2 galaxies, the values are between -0.92 and 0.53, while for the other set of galaxies: starburst, LIRGs and ULIRGs, the values are in the range of -1.63 and 0.5. \\

\begin{figure}[!t]
   \centering
   \includegraphics[width=\columnwidth]{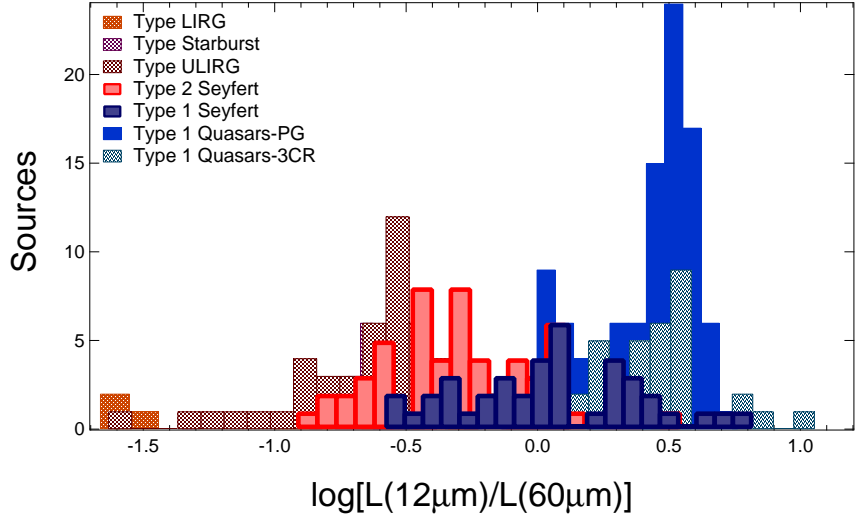} 
   \caption{\textrm{L}(12,60) histogram plot for all galaxies included in the analysis. The Seyfert 1 galaxies and quasars are well differentiated from the Seyfert 2, starburst, LIRG and ULIRG galaxies.}
   \label{fig:histogram}
\end{figure}

The separation between the \textit {Type 1} objects and the ULIRG-LIRG-Starburst  galaxies, also observed in  Seyfert 1 and 2 galaxies (Figure \ref{fig:histogram}), is consistent with the fact that emissions in the near and mid-infrared (NIR-MIR) spectra from the dust heated by the AGN, are dominant in  \textit {Type 1} objects. In order to explore this behavior, we applied the Kolmogorov-Smirnov (KS) test to \textit {Type 1} objects and  Seyfert 2 galaxies, and the results are shown in Table \ref{tableT1S2}. The P-values for L(12$\mu$m), L(60$\mu$m) and L(12$\mu$m)/L(60$\mu$m) reject the null hypothesis, showing significant  differences between the Mid-IR and Far-IR emissions. For the 7.7$\mu$m-PAH emissions,  the test also rejects the null hypothesis, i.e., there is no isotropy in the PAH emissions. Considering  the latter as a tracer of the existence of star formation activity, the KS test provides statistical validation for the observed differences in our sample. \\

 \begin{table}
\begin{small}
\begin{center}
\scalebox{0.8}{
\begin{tabular}{llllll}
\hline
\hline
{\bf K-S Test}	&	{\bf Log[L(12$\mu$m)]}	&	{\bf Log[L(60$\mu$m)]}	&	{\bf Log[L(12,60)} & {\bf Log[L(7.7$\mu$m)]} \\
\hline
$\alpha$	&	0.05	&	0.05	&	0.05	& 0.05  \\
{\it Type 1}	&	169	&	169	&	169	& 57 	\\
Sy2	&	53	&	53	&	53	& 47 	\\
D	&	0.681925	&	1	&	0.975215	& 0.574095 	\\
Cr\'itical	&	0.228102	&	0.228102	&	0.228102	 & 0.283724   \\
p-value	&	1.9598E-17	&	5.30404E-37	&	5.2752E-35  & 1.306E-8 	\\
\hline

Result	&	Reject	&	Reject	&  Reject & Reject 	\\
\hline
\hline
\end{tabular}}
\end{center}
\end{small}
\caption{\small K-S test done over  {\it Type 1} and Seyfert 2 galaxies.}
\label{tableT1S2}
\end{table}%

The anisotropies found in the KS test have two possible explanations: the PAH bands are diluted by intense X-ray emissions  (Yong et al. 2007) (a1), or  different stages of stellar activity are present in active galactic nuclei (a2). These two alternatives will be discussed in the following sections.\\

\begin{figure}[!t]
\begin{center}
   \includegraphics[width=12cm]{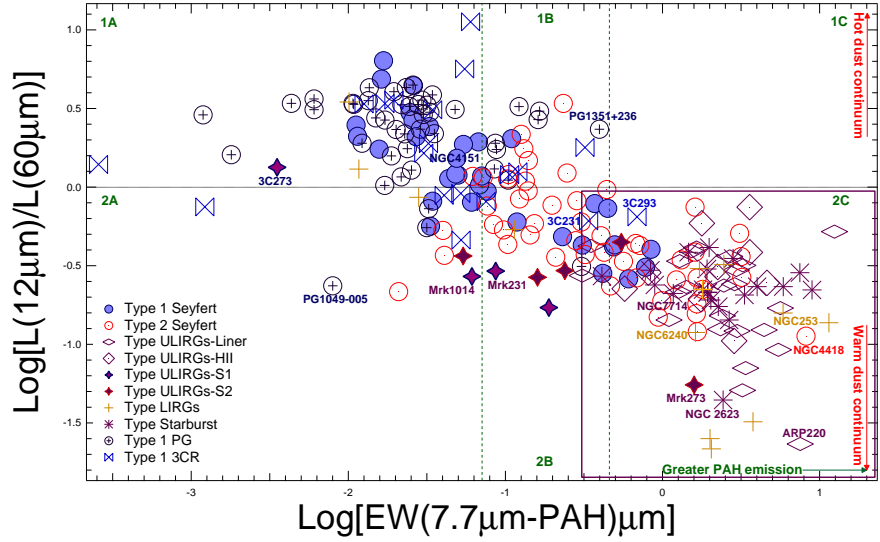}\vspace{10mm}
     \caption{Diagnostic plot of the L(12$\mu$m)/L(60$\mu$m) luminosity ratio as a function of the 7.7$\mu$m PAH equivalent width. The sample has a redshift $z$ between 0.000103, for the starburst IC342, and 0.55 for the radio source 3C330. The violet square box, at the right, defines the {\it ASFR} region.}
   \label{L12L60LogEW77}
 \end{center}  
\end{figure}

In order to separate the sample of galaxies by their degree of nuclear activity as well as the contribution associated with star formation, we selected the  $\textrm{L}(12,60)$ ratio and the 7.7$\mu$m PAH equivalent width for the construction of a new diagnostic diagram (Figure \ref{L12L60LogEW77}). \textit {Type 1} objects occupy the top half of the diagram, while the bottom half is mainly occupied by Seyfert 2, starburst galaxies, LIRGs and ULIRGs. This pattern is due to differences in the observed emissions at 12 $\mu$m and 60 $\mu$m. At higher values of $\textrm{L}(12,60)$, the dust is heated by the active nucleus and produces a warm black body emission \citep{pope08}, while at intermediate values of the ratio, the molecular clouds are heated by AGN and stellar activity. At lower values of $\textrm{L}(12,60)$ the star component heats the dust at temperatures between 30-50 K \citep{haas03}. The lowest $\textrm{L}(12,60)$ value, associated with the quasar PG1049-005, separates {\it Type 1} from the rest of the sample (Figure \ref{L12L60LogEW77}) and is equal to the separation value between Seyfert 1 and Seyfert 2 galaxies (-0.6), derived by \citet{mura00}. \\

The dotted vertical lines shown in figure \ref{L12L60LogEW77} correspond to the regions defined by \citet{spoo07}, but in our diagram, classes 1A-2A are occupied by unobscured AGNs, classes 1B-2B by composite (AGN/SB activity) galaxies and class 2C mainly by starburst galaxies. The square box located in the lower right quadrant delimits a particular region denominated by us as \textit {Active star formation region, ASFR}. In the \textit{ASFR} are located 78\% of LINERs and all starburst and ULIRGs galaxies, however 43\% of the Seyfert 2, and  11\% of Seyfert 1 samples are also present. \\

In the {\it ASFR} appear the LINERs: NGC1097 (not labeled in Figure \ref{L12L60LogEW77}) and NGC 253. NGC 1097 is a galaxy with a relatively young stellar population ($\sim 10^6$ years), at a distance less than 9 pc from the nucleus \citep{stor05, prie05}, that is the source of dust heating \citep{maso07}; NGC 253 is a galaxy in which a strong starburst and a weak AGN coexist \citep{mull10}.  The particular position of NGC 1097 and NGC 253 within the ULIRGs and starburst galaxies in the diagram, validates star formation as the cause of dust heating, and minimizes the role of the active nucleus. Also in this region appears the Seyfert 1 galaxy, NGC 3227 (just in the upper limit of the \textit{ASFR} frame); \citet{davi07} resolved the nuclear stellar distribution of this galaxy and found that within a few parsecs of the AGN there was an intense starburst, the most recent episode of which began $\sim$ 40 Myr ago but has now ceased. In the lower half of the {\it ASFR},  two interacting systems appear, NGC 2623 and ULIRG-LINER Arp 220. The first is a triple system with intense star formation, superluminal in the infrared, and extremely bright in radio \citep{read98}. The second is the most luminous object in the IRAS catalogue, also classified as `` heavily extinguished starburst " by \citet{hatz08}, and localized in the region 3B of the \citet{spoo07} diagram, an area in which absorption is the dominant factor. \\

The diagram in Figure \ref{L12L60LogEW77} is an original contribution to this field; it separates the sample in two groups of objects, one consisting of starburst-ULIRGs ({\it ASFR} region) and the other of {\it Type 1}-Seyfert 2 galaxies. In the ULIRG-starburst group, the stellar component is the factor responsible for warming the dust, which is reflected in the lower values for $\textrm{L}(12,60)$, and the values of PAH 7.7 $\mu$m ({\it ASFR} zone). The fact that some Seyfert galaxies with star formation appear in this area, confirms the usefulness of the diagram to validate the existence of star activity in AGNs. Additionally this diagram complements the \citet{spoo07} diagram with new data from IRAS observations and our measurements of PAH in 7.7$\mu$m.  Finally, our spectra extraction, using the PAHFIT tool and adjusting for every object the continuum, solves the overestimation in the 7.7 $\mu$m PAH equivalent width shown by \citet{clav00}.

\section {The FIR-Radio correlation}\label{corFIRRad}

The infrared emissions are largely due to the reemission of dust heated by massive stars, while supernova explosions of massive young stars are the source of radio emissions \citep{dejo85}. A quantifier of stellar activity is the slope of the correlation between radio emission and far-infrared emission or parameter {\it q} \citep{helo85,cond91}.

\begin{equation}
q\equiv  \log \frac{\textrm{F}_{\textrm{FIR}}/(3.75\times10^{12} \textrm{Hz})}{\textrm{F}_{\nu}(1.4\textrm{GHz})(\textrm{W m}^{-2} \textrm{Hz}^{-1})} ,
\label{parametroq}
\end{equation}

where $\textrm{F}_{\nu}$ (1.4GHz) is the density flux at 1.4 GHz, and $\textrm{F}_{\textrm{FIR}}$ is the flux in far infrared calculated as,

\begin{equation}
\textrm{FIR} \equiv 1.26 \times 10^{-14}(2.58 \textrm{F}_{60\mu m}+ \textrm{F}_{100\mu m}) \textrm{W m}^{-2}
\end{equation}

where F$_{60\mu \textrm{m}}$ and F$_{100\mu \textrm{m}}$ are the IRAS $60\mu \textrm{m}$ and $100\mu \textrm{m}$ band flux densities measured in Janskys (Jy). \\

At low IR luminosities, the average value of $q$ tends to increases from younger to older stellar populations \citep{cond91, xu94}. \citet{sand96} established, using IRAS observations, a $q$ value equal to 2.35 for objects with intense star formation, and \citet{saba08} used a FIR, F(1.4Ghz) correlation to identify galaxies with excess radio emission as AGN candidates.\\

\begin{figure}[!t]
\begin{center}
\includegraphics[width=12cm]{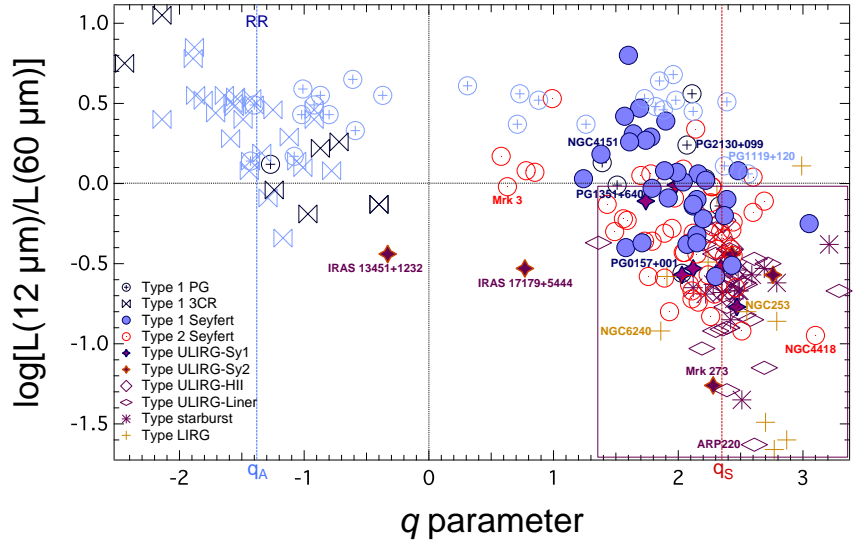}
\caption{\small  $\textrm{L}(12,60)$ indicator plotted against the $\it q$ parameter. The red dotted line marked with the label $\it q_S$ shows the value 2.35  \citep{sand96} for starburst galaxies. The light blue symbols represent quasars with upper limit estimations. The blue dotted line labeled $\it q_A$ ($-1.38$) represents the mean of all values  {\it q} $<$ 0. The graphic includes additional sources from \citet{schw06} and \citet{yong07}.}
\label{L12L60Paramq}
\end{center}
\end{figure}

 In this context, the parameter $q$ is also a discriminator between sources that have strong star formation and those (AGNs) in which the radio luminosity is greater than the value predicted by the radio-FIR correlation \citep{yun01}. Therefore, we conducted a review of the $q$ parameter for our sample. Figure \ref{L12L60Paramq} shows the graph of the $\textrm{L}(12,60)$ indicator in function of the parameter $q$. Following the distribution of the sample along the vertical axis a frame is drawn, indicating the  \textit {Active star formation region} area. The dotted lines at the origin separate the sample by the dominant emission. On the vertical axis, the sources are discriminated according to emission intensity, between mid-infrared and cold dust (60 $\mu$m). On the horizontal axis, the objects are separated into those with strong radio emission (left) and those with dominance in the far-infrared emission (right). The dashed line labeled q${_\textrm{s}}$ shows the value determined by \citet{sand96}. \\

In Figure \ref{L12L60Paramq}, three QSOs: PG 1351+640, PG 1119+120 and PG2130+099 are in the region with positive parameter $q$. \citet{farr09} also present these three sources and suggest that they represent the last part of an evolutionary sequence, post-IR, where the contribution of star formation falls to mid-infrared, at the expense of increasing the contribution of the AGN. However, as the diagram also shows, a large sample of galaxies are in the region associated with parameter values {\it q} $<$ 0, here called: {\it Radio Region, RR}. In radio galaxies, the accretion of massive black holes located in giant ellipticals is less than $10^{-3}$ M$_\odot$/year, and this condition restricts star formation activity \citep{dick11}. In addition, radio jets ionize the gas, cooling it and reducing the activity of star formation and dust heating \citep{ramo09}. This graph is particularly suitable for discriminating highly noisy radio galaxies with low density material from those with intense FIR emissions associated with dense clouds. The sources located in the radio region RR are not directly linked to evolutionary scenarios presented in the literature; whether or not they are a final stage of an evolutionary scenario is an open question. However, the graph confirms that they are active nuclei with low stellar activity.\\

\section{Conclusions}
The diagnostic diagram, in which $\textrm{L}(12,60)$ is plotted against  the equivalent width of the 7.7$\mu$m PAH emission, is an original tool for discriminating galaxies by strength and source of activity (AGN or star formation). The ULIRG-starburst galaxies group in the {\it ASFR} region, verifying that the stellar component is the dominant factor responsible for the heating of dust, while, for the remaining galaxies, the AGN component is dominant.

The galaxies sampled in this study overlap with the sampling of previous authors  \citep{stor01, haas03, koul06, farr09}, even though each study used different variables. Nevertheless, in all of our studies, including this one, the distribution of galaxies in the parameter space is similar and this demonstrates the discriminatory power of our method. The relative position of NGC 253, a weak AGN, on the right side, NGC 1027, a LINER, in the middle and NGC3227, a Seyfert 1, on the left side of the {\it ASFR} region, validates the coexistence of  star formation and an active nucleus. \\

The diagnostic diagram of $\textrm{L}(12,60)$ vs parameter $q$, validates the relationship between far-infrared emission and the emission at 1.4 GHz (associated with supernova-type events) that holds for galaxy type starbursts, ULIRGs and LIRGs. The strong 1.4 GHz emission from an active nucleus defines a new region with negative $q$ values  (region {\it RR}) where 3CR radio sources, some of the quasar (PG) and a ULIRG-SY2 are located. There, the stellar activity for radio galaxies is lower. This diagram, like the previous one, is useful in the identification of objects, according to their level of nuclear activity, in large observational surveys.\\

The scenarios (a1) and (a2), proposed in section \ref{MidFIR}, coexist in the diagnostic diagrams. The dominant activity in a galaxy determines the predominance of one or the other scenario and its relative location in one of the evolutionary sequences suggested by \citet{stor01}, \citet{haas03}, \citet{farr09}. \\

The authors wish to thank the anonymous reviewer for his/her very careful reading of the manuscript and very important and fruitful comments and suggestions. We also thank Alberto Rodr\'iguez Ardila for useful and fundamental discussions, and Lauren Raz for her English style suggestions.

\begin{center}

\begin{table}[htbp]
\caption{Data from literature and our estimations}
\begin{center}
\scalebox{0.54}{
\begin{tabular}{lcclcc}
\hline \hline 
\textbf{Seyfert 1} & Log${(\textrm{L}12/\textrm{L}60)}$ & L$_{7.7 \mu \textrm{m}} $ ($\textrm{W}/\textrm{m}^{2}$) & EQW$_{7.7\mu \textrm{m}}$ ($\mu \textrm{m}$) & Flux(FIR)($\textrm{W}\textrm{m}^{-2}$) &F$_{\textrm{1.4GHz}}$ ($\textrm{W}\textrm{m}^{-2}\textrm{Hz}^{-1}$)\\ 
\hline  
NGC863(Mrk590) & 0.29 & 1.911E+38 & 0.067 & 3.43E-14 &1.52E-28 \\ 
NGC3786(Mrk744) & -0.38 & 1.326E+38 & 0.489 & 7.68E-14 & 1.75E-28 \\ 
NGC4235 & 0.31 & 1.979E+37 & 0.110 & 1.86E-14 &	1.15E-28\\ 
NGC4253(Mrk766) & -0.32 & 3.484E+38 & 0.231 & 1.90E-13 &	3.58E-28 \\ 
NGC5548 & 0.27 & 1.185E+38 & 0.054 & 5.51E-14 & 2.65E-28\\ 
Mrk817 & -0.10 & 5.176E+38 & 0.061 & 9.75E-14 &	1.05E-28 \\ 
NGC7469 & -0.58 & 3.497E+39 & 0.609 & 1.28E-12	& 1.70E-27 \\ 
Mrk530(NGC7603) & -0.03 & 1.020E+39 & 0.077 & 5.33E-14	& 2.29E-28 \\ 
NGC931(Mrk1040) & 0.08 & 2.022E+38 & 0.050 & 1.41E-13	& 1.25E-28 \\ 
Iras F03450+0055 & 0.47 & 2.177E+38& 0.025 & 5.61E-14 &	3.05E-28 \\ 
MCG-5-13-17 & -0.10 & 2.085E+38 & 0.371 & 7.06E-14	& 1.33E-28 \\ 
Mrk79 & 0.01 & 3.876E+38 & 0.068 & 7.85E-14	& 1.93E-28 \\ 
NGC2639 & -0.40 & 2.905E+38 & 0.845 & 1.54E-13	& 1.08E-27 \\ 
NGC2992 & -0.37 & 2.870E+38 & 0.496 & 4.05E-13 &	2.13E-27 \\ 
UGC7064 & -0.51 & 1.134E+39 & 0.781 & 1.60E-13	& 1.57E-28 \\ 
MCG-2-33-34(NGC4748) & -0.14 & 2.118E+38 & 0.448 & 6.57E-14	& 1.32E-28 \\ 
IC4329A & 0.42 & 2.541E+38 & 0.026 & 8.69E-14	& 6.24E-28 \\ 
Mrk509 & 0.07 & 7.588E+38 & 0.071 & 6.34E-14	& 1.75E-28 \\ 
Mrk478 & 0.02 & 1.127E+39 & 0.052 & 3.01E-14 &	4.89E-29 \\ 
NGC3227 & -0.37 & 4.897E+37 & 0.309 & 4.87E-13	& 9.17E-28 \\ 
Ark120 & 0.39 & 2.417E+38 & 0.011 & 3.45E-14	& 1.16E-28 \\ 
NGC4051 & -0.22 & 1.094E+37 & 0.119 & 5.33E-13	& 8.87E-28 \\ 
NGC4593 & -0.25 & 4.028E+37 & 0.033 & 1.74E-13	& 4.13E-29 \\ 
Mrk279 & -0.09 & 2.206E+38 & 0.034 & 6.87E-14	& 2.18E-28 \\ 
NGC526a & 0.80 & 6.143E+37 & 0.017 & 1.83E-14	& 1.23E-28 \\ 
NGC3516 & 0.08 & 4.459E+37 & 0.048 & 8.57E-14	& 2.94E-28 \\ 
NGC4151	& 0,18 & 3.627E+37 & 0.049 &3.22E-13 	& 3.60E-27\\
\hline \hline 
\textbf{Seyfert 2} & Log${(\textrm{L}12/\textrm{L}60)}$ & L$_{7.7 \mu \textrm{m}} $ ($\textrm{W}/\textrm{m}^{2}$) & EQW$_{7.7\mu \textrm{m}}$ ($\mu \textrm{m}$) & Flux(FIR)($\textrm{W}\textrm{m}^{-2}$) &F$_{\textrm{1.4GHz}}$ ($\textrm{W}\textrm{m}^{-2}\textrm{Hz}^{-1}$) \\ 
\hline
Mrk334 & -0.58 & 2.357E+39 & 0.6815 & 1.96E-13 &	2.62E-28\\ 
Mrk993 & 0.34 & 1.593E+37 & 0.126 & 2.64E-14 &	5.07E-29 \\ 
Mrk573 & 0.06 & 2.489E+38 & 0.07205 & 5.13E-14	& 2.26E-28 \\ 
NGC1144 & -0.58 & 9.577E+39 & 1.66 & 3.15E-13	& 1.45E-27 \\ 
NGC4388 & -0.31 & 7.267E+38 & 0.406 & 5.60E-13	& 1.12E-27 \\ 
NGC5252 & -0.67 & 2.135E+37 & 0.0209 & 7.44E-14	 & 1.53E-28 \\ 
NGC5256(Mrk266SW) & -0.80 & 8.364E+39 & 1.64 & 3.78E-13	& 1.19E-27 \\ 
NGC5347 & 0.04 & 9.908E+37 & 0.134 & 7.94E-14	& 5.26E-29 \\ 
NGC5695 & -0.02 & 5.655E+37 & 0.442 & 4.11E-14 &	5.92E-29 \\ 
NGC5929 & -0.63 & 2.368E+37 & 0.4665 & 4.70E-13	 & 1.02E-27 \\ 
NGC7674 & -0.22 & 5.823E+39 & 0.304 & 2.84E-13	& 2.08E-27 \\ 
NGC7682 & 0.53 & 2.839E+37 & 0.2335 & 2.04E-14	& 5.62E-28 \\ 
Mrk938 & -0.92 & 6.217E+39 & 1.665 & 7.56E-13	& 6.29E-28 \\ 
NGC262(Mrk348) & 0.08 & 3.701E+38 & 0.1045 & 6.15E-14	& 2.75E-27 \\ 
NGC513 & -0.36 & 1.225E+39 & 0.675 & 1.14E-13 &	4.97E-28 \\ 
F01475-0740 & 0.17 & 1.984E+38 & 0.14 & 4.23E-14	& 3.00E-27 \\ 
NGC1125 & -0.59 & 4.087E+38 & 1.225 & 1.57E-13 &	5.45E-28 \\ 
NGC1320(Mrk607) & -0.11 & 3.196E+38 & 0.198 & 1.05E-13 &	5.64E-29 \\ 
F04385-0828 & -0.12 & 6.072E+38 & 0.07665 & 1.22E-13	& 1.79E-28 \\ 
NGC1667 & -0.44 & 3.594E+39 & 3.16 & 3.79E-13	& 7.12E-28 \\ 
NGC3660 & -0.29 & 6.660E+38 & 3.095 & 1.18E-13	& 1.16E-28 \\ 
NGC4501-S1-2 & -0.43 & 3.549E+38 & 0.322 & 1.13E-12 &	2.79E-27 \\ 
NGC4968 & -0.09 & 3.625E+38 & 0.285 & 1.14E-13 & 3.24E-28 \\ 
MCG-3-34-64 & -0.13 & 6.764E+38 & 1.61 & 2.61E-13 &	2.58E-27 \\ 
NGC5135 & -0.72 & 3.521E+39 & 0.989 & 9.10E-13	& 1.88E-27 \\ 
MCG-2-40-4 & -0.27 & 1.714E+39 & 0.09725 & 2.01E-13	& 2.82E-28 \\ 
NGC7172 & -0.41 & 5.551E+38 & 0.4245 & 3.41E-13 & 	3.46E-28 \\ 
MCG-3-58-7 & -0.24 & 1.642E+39 & 0.0839 & 1.21E-13 &	1.17E-28 \\ 
IC3639 & -0.37 & 1.361E+39 & 0.717 & 3.79E-13 &	8.26E-28 \\ 
Mrk34 & -0.36 & 1.003E+39 & 0.1035 & 3.64E-14 &	1.79E-28 \\ 
Mrk78 & -0.23 & 9.051E+38 & 0.153 & 5.03E-14 &	3.43E-28 \\ 
Mrk463 & 0.07 & 4.717E+39 & 0.06185 & 9.51E-14 &	3.58E-27 \\ 
Mrk477 & -0.3 & 8.556E+38 & 0.1435 & 6.59E-14	& 5.67E-28 \\ 
NGC1068 & 0.05 & 2.881E+39 & 0.1034 & 8.55E-12	 & 4.56E-26 \\ 
NGC5033 & -0.46 & 2.272E+38 & 1.315 & 1.00E-12	& 1.14E-27 \\ 
IRAS 22377+0747 & 0.09 & 5.079E+38 & 0.2465 & 5.81E-14	& 1.42E-28 \\ 
NGC4579 & -0.43 & 1.662E+37 & 0.04085 & 3.82E-13 &	9.10E-28 \\ 
NGC7314 & -0.45 & 2.013E+37 & 0.209 & 3.00E-13 & 	2.91E-28 \\ 
NGC1097 & -0.65 & 1.187E+39 & 1.8 & 2.52E-12 &	2.34E-27 \\ 
NGC1386 & -0.34 & 6.069E+37 & 0.2805 & 2.97E-13 & 	3.49E-28 \\ 
Mrk3 & -0.02 & 4.356E+38 & 0.1395 & 1.65E-13	& 1.03E-26\\ 
NGC3982 & -0.41 & 2.164E+38 & 1.61 & 4.05E-13 & 	5.30E-28 \\ 
NGC4507 & -0.28 & 2.017E+38 & 0.04 & 2.08E-13	& 6.21E-28 \\ 
NGC5728 & -0.83 & 3.584E+38 & 0.9385 & 4.50E-13 &	6.58E-28 \\ 
NGC5953 & -0.57 & 9.253E+38 & 3.195 & 5.64E-13 &	8.59E-28 \\ 
NGC7592 & -0.74 & 1.109E+40 & 1.655 & 3.82E-13	& 7.06E-28 \\ 
NGC4418 (NGC4355) & -0,95 & 1,182E+39  &  8.19&	1.83E-12	& 3.85E-28\\
\hline 
\end{tabular}}
\end{center}
\label{datasource}
\end{table}
\end{center}

\begin{center}
\begin{table}[htbp]
\begin{center}
\scalebox{0.54}{
\begin{tabular}{lcclcc}
\hline \hline 
\textbf{LIRGs} & Log${(\textrm{L}12/\textrm{L}60)}$ & L$_{7.7 \mu \textrm{m}} $ ($\textrm{W}/\textrm{m}^{2}$) & EQW$_{7.7\mu \textrm{m}}$ ($\mu \textrm{m}$) & Flux(FIR)($\textrm{W}\textrm{m}^{-2}$) &F$_{\textrm{1.4GHz}}$ ($\textrm{W}\textrm{m}^{-2}\textrm{Hz}^{-1}$) \\
\hline  
NGC6240 & -0.92 & 1.467E+40 & 1.62 & 1.10E-12 &	4.01E-27 \\ 
Iras23060+0505 & -0.07 & 1.266E+40 & 0.02805 & 4.78E-14	& 6.30E-29 \\ 
Iras20460+1925 & 0.11 & 5.326E+39 & 0.01167 & 6.33E-13	& 1.74E-28 \\ 
NGC253 & -0.80 & 1.304E+38 & 5.85 & 3.78E-11 &	2.81E-26 \\ 
NGC828 & -0.49 & 1.430E+40 & 2.455 & 6.42E-13	& 9.76E-28 \\ 
Iras15250+3609 & -0.86 & 2.200E+40 & 11.455 & 3.11E-13 &	1.36E-28 \\ 
Iras17208-0014 & -1.49 & 2.084E+40 & 3.765 & 1.45E-12	 & 7.69E-28 \\ 
Iras13126+2452 & -1.60 & 2.889E+38 & 2.005 & 8.10E-13	 & 2.89E-28 \\ 
CGCG1510.8+0725 & -1.66 & 8.913E+38 & 2.045 & 1.05E-12	& 4.79E-28 \\ 
\hline 
\end{tabular}}
\end{center}
\label{}
\end{table}

\end{center}

\begin{center}
\begin{table}[htbp]
\begin{center}
\scalebox{0.54}{
\begin{tabular}{lcclcc}
\hline \hline
\textbf{ULIRG-LINER}& Log${(\textrm{L}12/\textrm{L}60)}$ & L$_{7.7 \mu \textrm{m}} $ ($\textrm{W}/\textrm{m}^{2}$) & EQW$_{7.7\mu \textrm{m}}$ ($\mu \textrm{m}$) & Flux(FIR)($\textrm{W}\textrm{m}^{-2}$) &F$_{\textrm{1.4GHz}}$ ($\textrm{W}\textrm{m}^{-2}\textrm{Hz}^{-1}$) \\ 
\hline
IRAS00188-0856 & -0.64 & 2.592E+40 & 1.416 & 1.27E-13 &	1.48E-28 \\ 
IRAS03250+1606 & -0.44 & 1.842E+40 & 2.68 & 6.72E-14 &	8.93E-29 \\ 
IRAS08572+3915 & -0.67 & 6.346E+40 & 1.264 & 2.99E-13	& 4.09E-29 \\ 
IRAS09039+0503 & -0.63 & 1.563E+40 & 2.455 & 7.41E-14 &	5.83E-29 \\ 
IRAS09116+0334 & -0.38 & 8.951E+39 & 1.6455 & 5.84E-14 &	9.87E-29 \\ 
IRAS09539+0857 & -0.28 & 1.175E+41 & 12.425 & 5.99E-14 &	7.99E-29 \\ 
IRAS10378+1108(9) & -0.62 & 1.868E+40 & 2.875 & 9.71E-14 &	7.89E-29 \\ 
IRAS10485-1447 & -0.5 & 1.828E+40 & 3.245 & 7.72E-14	& 3.90E-29 \\ 
IRAS10494+4424 & -0.77 & 1.605E+40 & 5.62 & 1.83E-13 &	1.99E-28 \\ 
IRAS11095-0238 & -1.03 & 5.148E+40 & 5.46 & 1.38E-13	 & 2.36E-28 \\ 
IRAS12112+0305 & -1.15 & 1.583E+40 & 3.38 & 4.02E-13 &	2.19E-28 \\ 
IRAS12127-1412 & -0.37 & 2.240E+40 & 0.3505 & 6.43E-14	& 7.46E-28 \\ 
IRAS12359-0725 & -0.47 & 1.066E+40 & 1.285 & 5.73E-14 &	3.76E-29 \\ 
IRAS14252-1550 & -0.41 & 1.252E+40 & 1.535 & 6.08E-14	& 6.30E-29 \\ 
IRAS14348-1447 & -1.29 & 2.584E+40 & 3.225 & 3.12E-13	& 3.37E-28 \\ 
IRAS15327+2340(Arp220) & -1.63 & 8.163E+39 & 7.52 & 4.78E-12	& 3.11E-27 \\ 
IRAS16090-0139 & -0.9 & 5.526E+40& 2.34 & 1.79E-13	& 1.96E-28 \\ 
IRAS16487+5447 & -0.92 & 6.271E+39 & 2.98 & 1.32E-13	& 1.79E-28 \\ 
IRAS17028+5817 & -0.91 & 1.192E+40 & 4.435 & 1.28E-13 &	1.48E-28 \\ 
IRAS17044+6720 & -0.56 & 2.367E+40 & 0.308 & 5.40E-14 &	4.70E-29 \\ 
IRAS21329-2346 & -0.82 & 1.598E+40 & 2.405 & 8.16E-14	& 6.58E-29 \\ 
\hline\hline
\textbf{ULIRG-HII} & Log${(\textrm{L}12/\textrm{L}60)}$ & L$_{7.7 \mu \textrm{m}} $ ($\textrm{W}/\textrm{m}^{2}$) & EQW$_{7.7\mu \textrm{m}}$ ($\mu \textrm{m}$) & Flux(FIR)($\textrm{W}\textrm{m}^{-2}$) &F$_{\textrm{1.4GHz}}$ ($\textrm{W}\textrm{m}^{-2}\textrm{Hz}^{-1}$) \\
\hline 
IRAS11387+4116 & -0.23 & 1.237E+40 & 1.78 & 5.22E-14	 & 6.11E-29 \\ 
IRAS11506+1331 & -0.71 & 3.430E+40 & 1.755 & 1.26E-13 &	1.27E-28 \\ 
IRAS13539+2920 & -0.61 & 1.698E+40 & 4.62 & 9.39E-14	& 1.09E-28 \\ 
IRAS14060+2919 & -0.51 & 2.337E+40 & 2.245 & 8.28E-14 &	8.74E-29 \\ 
IRAS15206+3342 & -0.65 & 2.699E+40 & 0.574 & 8.14E-14 &	1.01E-28 \\ 
IRAS15225+2350 & -0.57 & 1.852E+40 & 2.235 & 6.09E-14 &	6.20E-29 \\ 
IRAS16474+3430 & -0.54 & 1.019E+40 & 0.8125 & 110E-13 &	1.04E-28 \\ 
IRAS21208-0519 & -0.41 & 1.144E+40 & 3.595 & 5.90E-14	& 6.01E-29 \\ 
\hline \hline
\textbf{ULIRG-Seyfert 1} & Log${(\textrm{L}12/\textrm{L}60)}$ & L$_{7.7 \mu \textrm{m}} $ ($\textrm{W}/\textrm{m}^{2}$) & EQW$_{7.7\mu \textrm{m}}$ ($\mu \textrm{m}$) & Flux(FIR)($\textrm{W}\textrm{m}^{-2}$) &F$_{\textrm{1.4GHz}}$ ($\textrm{W}\textrm{m}^{-2}\textrm{Hz}^{-1}$) \\
\hline 
IRAS01572+0009(Mrk1014) & -0.57 & 8.698E+39 & 0.0612 & 9.94E-14	& 2.46E-28 \\ 
IRAS12265+0219(3C273) & 0.13 & 1.363E+39 & 0.00352 & 1.03E-13 &	5.17E-25 \\ 
IRAS12540+5708(Mrk231) & -0.53 & 1.436E+40 & 0.08665 & 1.42E-12 &	2.90E-27 \\ 
IRAS15462-0450 & -0.77 & 1.058E+40 & 0.189 & 1.33E-13 &	1.20E-28 \\ 
\hline\hline
\textbf{ULIRG-Seyfert 2} & Log${(\textrm{L}12/\textrm{L}60)}$ & L$_{7.7 \mu \textrm{m}} $ ($\textrm{W}/\textrm{m}^{2}$) & EQW$_{7.7\mu \textrm{m}}$ ($\mu \textrm{m}$) & Flux(FIR)($\textrm{W}\textrm{m}^{-2}$) &F$_{\textrm{1.4GHz}}$ ($\textrm{W}\textrm{m}^{-2}\textrm{Hz}^{-1}$) \\ 
\hline
IRAS05189-2524 & -0.57 & 8.346E+39 & 0.16 & 5.88E-13	 & 2.71E-28 \\ 
IRAS08559+1053 & -0.35 & 2.513E+40 & 0.5465 & 6.10E-14 &	1.15E-28 \\ 
IRAS13428+5608(Mrk273) & -1.26 & 1.101E+40 & 1.59 & 9.76E-13 &	1.36E-27 \\ 
IRAS13451+1232(PKS1345+12) & -0.44 & 3.343E+39 & 0.0538 & 8.84E-14	& 5.07E-26 \\ 
IRAS17179+5444 & -0.53 & 9.085E+39 & 0.239 & 6.83E-14 &	3.12E-27 \\ 
\hline
\end{tabular}}
\end{center}
\label{}
\end{table}

\end{center}

\newpage

\begin{center}
\begin{table}[htbp]
\begin{center}
\scalebox{0.54}{
\begin{tabular}{lcclcc}
\hline \hline
\textbf{Starburst} & Log${(\textrm{L}12/\textrm{L}60)}$ & L$_{7.7 \mu \textrm{m}} $ ($\textrm{W}/\textrm{m}^{2}$) & EQW$_{7.7\mu \textrm{m}}$ ($\mu \textrm{m}$) & Flux(FIR)($\textrm{W}\textrm{m}^{-2}$) &F$_{\textrm{1.4GHz}}$ ($\textrm{W}\textrm{m}^{-2}\textrm{Hz}^{-1}$) \\
\hline 
IC342 & -0.38 & 1.488E+36 & 1.98 & 1.08E-11	& 1.79E-27 \\ 
Mrk52 & -0.53 & 1.166E+38 & 0.8375 & 2.25E-13 &	1.23E-28 \\ 
NGC0520 & -0.85 & 3.895E+39 & 2.655 & 1.62E-12	& 1.66E-27 \\ 
NGC1222 & -0.72 & 4.933E+38 & 2.03 & 6.19E-13 &	5.80E-28 \\ 
NGC1365 & -0.57 & 5.022E+38 & 0.4115 & 5.15E-12	 & 3.53E-27\\ 
NGC2146 & -0.63 & 7.437E+38 & 4 & 7.21E-12	& 1.01E-26 \\ 
NGC2623 & -1.35 & 4.323E+39 & 2.43 & 1.10E-12 &	8.99E-28 \\ 
NGC3310 & -0.65 & 9.577E+37 & 1.81 & 1.68E-12 &	3.41E-27 \\ 
NGC3556 & -0.45 & 2.910E+37 & 2.815 & 2.03E-12	& 2.03E-27 \\ 
NGC3628 & -0.54 & 2.434E+38 & 7.485 & 3.11E-12 &	2.73E-27\\ 
NGC4194 & -0.67 & 1.484E+39 & 1.935 & 1.07E-12	& 9.46E-28\\ 
NGC4676 & -0.69 & 3.494E+39 & 3.325 & 1.52E-13	& 2.46E-28 \\ 
NGC4818 & -0.62 & 1.970E+38 & 0.892 & 9.89E-13	& 4.28E-28 \\ 
NGC7252 & -0.52 & 3.191E+39 & 1.92 & 2.18E-13 & 	2.31E-28 \\ 
NGC7714 & -0.68 & 7.515E+38 & 1.17 & 5.17E-13 &	 6.18E-28 \\ 
NGC1614 & -0.67 & 4.715E+39 & 1.945 & 1.48E-12	& 1.29E-27\\ 
\hline
\end{tabular}}
\end{center}
\label{}
\end{table}

\end{center}

\begin{center}
\begin{table}[htbp]
\begin{center}
\scalebox{0.54}{
\begin{tabular}{lcclcc}
\hline \hline
\textbf{QSOS-QUASARs PG}& Log${(\textrm{L}12/\textrm{L}60)}$ & L$_{7.7 \mu \textrm{m}} $ ($\textrm{W}/\textrm{m}^{2}$) & EQW$_{7.7\mu \textrm{m}}$ ($\mu \textrm{m}$) & Flux(FIR)($\textrm{W}\textrm{m}^{-2}$) &F$_{\textrm{1.4GHz}}$ ($\textrm{W}\textrm{m}^{-2}\textrm{Hz}^{-1}$) \\ 
\hline
PG1229+204(Mrk771) & 0.34 & 3.182E+38 & 0.02285 & \textit{1.18E-14} &	\textit{2.63E-29} \\ 
PG1302-102 & 0.24 & 1.590E+40& 0.08765 &\textit{1.55E-14}	& \textit{6.68E-27} \\ 
PG1426+015 & 0.28 & 6.972E+38 & 0.01223 & \textit{1.53E-14}	& \textit{2.54E-29 }\\ 
PG2349-014 & 0.28 & 3.045E+39 & 0.0862 & \textit{1.25E-14}	& \textit{1.51E-26} \\ 
PG0007+106 & 0.37 & 9.131E+38 & 0.02015 & \textit{1.76E-14}	& \textit{9.19E-28} \\ 
PG1115+407 & 0.51 & 4.891E+39 & 0.122 &\textit{ 8.92E-15}	& \textit{9.68E-30} \\ 
PG1119+120 & 0.11 & 4.083E+38 & 0.02525 & \textit{2.49E-14} &	\textit{2.82E-29} \\ 
PG1351+640 & -0.01 & 3.979E+39 & 0.3965 & 3.76E-14	& 3.07E-28 \\ 
PG1404+226 & 0.52 & 1.449E+38 & 0.0219 & \textit{9.38E-15}	& \textit{2.63E-29} \\ 
PG1448+273 & 0.48 & 4.580E+38 & 0.033065 & \textit{6.98E-15} & 	\textit{2.82E-29} \\ 
PG1534+580 & 0.46 & 1.280E+38 & 0.03 & \textit{1.29E-14} &	\textit{4.51E-29} \\ 
PG1612+261 & 0.37 & 1.496E+39 & 0.0287 & \textit{1.14E-14 } &	\textit{1.68E-28} \\ 
PG1613+658(Mrk876) & -0.14 & 4.017E+39 & 0.03245 & 2.96E-14	& 3.66E-29 \\ 
PG2130+099 & 0.24 & 1.102E+39 & 0.02375 & 2.50E-14	& 5.64E-29 \\ 
\hline \hline
\textbf{QSOS-QUASAR 3CR} & Log${(\textrm{L}12/\textrm{L}60)}$ & L$_{7.7 \mu \textrm{m}} $ ($\textrm{W}/\textrm{m}^{2}$) & EQW$_{7.7\mu \textrm{m}}$ ($\mu \textrm{m}$) & Flux(FIR)($\textrm{W}\textrm{m}^{-2}$) &F$_{\textrm{1.4GHz}}$ ($\textrm{W}\textrm{m}^{-2}\textrm{Hz}^{-1}$) \\ 
\hline
3C48 & -0.34 & 4.006E+40 & 0.0522 & \textit{3.85E-14}	& \textit{1.51E-25} \\ 
3C65 & 0.55 & 1.053E+41 & 0.10945 & \textit{2.52E-15}	& \textit{2.92E-26} \\ 
3C79 & -0.09 & 7.999E+39 & 0.07533 & \textit{8.84E-15 }&	\textit{4.57E-26} \\ 
3C84 & -0.13 & 9.167E+36 & 0.00122 & 3.20E-13 &	2.15E-25 \\ 
3C249.1 & 0.14 & 3.180E+37 & 0.000257 & \textit{3.05E-15}	& \textit{2.20E-26} \\ 
3C270 & 1.05 & 4.527E+36 & 0.06 & 4.49E-15	& 1.66E-25 \\ 
3C272.1 & 0.22 & 8.945E+35 & 0.0302 & 2.87E-14 & 	5.70E-26 \\ 
3C274 & 0.75 & 2.825E+36 & 0.055 & 1.78E-14 & 	1.30E-24 \\ 
3C293 & -0.19 & 2.055E+39 & 0.688 & 1.56E-14 &	3.84E-26 \\ 
3C303.1 & 0.1 & 1.317E+39 & 0.12005 &\textit{ 6.53E-15}	& \textit{1.77E-26} \\ 
3C323.1 & 0.49 & 2.553E+39 & 0.0344 & \textit{3.40E-15} & 	\textit{2.25E-26} \\ 
3C330 & 0.08 & 7.048E+39 & 0.1064 & \textit{8.91E-15}	& \textit{6.57E-26} \\  
3C381 & 0.28 & 1.071E+39 & 0.03238 & \textit{3.66E-15} &	\textit{3.79E-26 }\\ 
3C382 & 0.53 & 1.928E+38 & 0.0126 & \textit{7.32E-15} &	\textit{5.16E-26} \\ 
3C386 & 0.55 & 1.637E+36 & 0.0195 & \textit{3.34E-15} &	\textit{6.54E-26} \\ 
3C445 & 0.53 & 3.817E+38 & 0.0171 & \textit{2.29E-14}	& \textit{5.79E-26} \\ 
\hline
\end{tabular}}
\end{center}
\textrm{{\tiny The data in italic text show upper limits. }}
\label{pg3cr}
\end{table}

\end{center}




\begin{thebibliography}
\bibitem[Allamandola et al., 1989]{alla89} Allamandola, L. J., Tielens, A.G.G.M and Barker, J. R. 1989. ApJS, 71, 733.
\bibitem[Allamandola et al., 1999]{alla99} Allamandola, L. J.,  Hudgins, D. M. and Sandford, S. A. 1999. ApJ, 511, L115.
\bibitem[Antonucci \& Miller., 1985]{anto85} Antonucci, R. \& Miller, J. 1985. ApJ, 297, 621.
\bibitem[Antonucci, 1993]{anto93} Antonucci, R. 1993.  ARA\&A, 31, 473.
\bibitem[Brand et al., 2006]{bran06}Brandl, B.,  et al. 2006, ApJ, 653, 1129.
\bibitem[Cepa, 2009]{cepa09}Cepa Jordi.  2009. {\it The Emission-Line Universe}, Cambridge University Press.  XVIII Canary Islands Winter School of Astrophysics, pp. 138-182.
\bibitem[Cid Fernandez et al., 2001]{cidf01}Cid Fernandes, R, Heckman, T, Schmitt, H, Gonz\'alez Delgado, R. M, Storchi-Bergmann, T. 2001. ApJ, 558, 81.
\bibitem[Clavel et al., 2000]{clav00} Clavel J, Shulz, B., Altieri, B., Barr, P., Claes, P., Heras, A., leech, K., Metcalfe, L., and Salama, A. 2000, A\&A, 357, 839.
\bibitem[Condon, 1991]{cond91}Condon, J. 1991. Annu. Rev. Astrom. Astrophys, 30, 575
\bibitem[Condon, 1998]{cond98} Condon, J.J., Cotton, W. D., Greisen, E. W., Yin, Q. F., Perley, R. A., Taylor, G. B., \& Broderick, J. J. 1998, AJ, 115, 1693
\bibitem[Davies et al., 2007]{davi07}Davies, R. I., Mueller S\'anchez, F., Genzel, R., Tacconi, L. J., Hicks, E. K. S., Friedrich, S., \& Sternberg, A. 2007, ApJ, 671, 1388
\bibitem[De Jong et al., 1985]{dejo85}De Jong, T., Klein, U., Wielebinsky, R., \& Wunderlich, E. 1985, A\&A, 147, L6.
\bibitem[Aleksandar et al., 2012]{alex12}Diamond-Stanic, Aleksandar M, Rieke, George H. 2012, ApJ, 746, 168.
\bibitem[Dicken et al., 2011]{dick11} Dicken, D, Tadhunter, C, Axon, D, Morganti, R, Robinson, A, Kouwenhoven, M. BN, Spoon, H, Kharb, P, Inskip, K. J, Holt, J, Almeida, C. R, Nesvadba, N. PH. 2011. arXiv 1111.4476v2.
\bibitem[Fanaroff \& Riley, 1974]{fana74} Fanaroff, B. L.\& Riley, J. M. 1974, MNRAS, 167, 31.
\bibitem[Farrah et al., 2008]{farr08} Farrah, D., Lonsdale, C. J., Weedman, D. W., Spoon, H. WW., Rowan-Robinson, M., Polletta, M., Oliver, S., Houck, J. R., and Smith, H. E. 2008, ApJ, 677, 957.
\bibitem[Farrah et al., 2009]{farr09} Farrah, D, Connolly, B, Connolly, N, Spoon, H. W. W, Oliver, S, Prosper, H. B, Armus, L, Houck, J. R, Liddle, A. R, Desai, V. 2009. ApJ, 700, 395.
\bibitem[Feigelson, 2001]{feig01}Feigelson, E. D. 2001. ASP Conference Series. 2001, 234, 131.
\bibitem[Feigelson et al., 2005]{feig05}Feigelson, E. D., Getman, K., Townsley, L., Garmire, G., Preibisch, T., Grosso, N., Montmerle, T., Muench, A., McCaughrean, Mark. 2005. ApJSS,  160, 379.
\bibitem[Freudling et al., 2003]{freu03} Freudling, W., Siebenmorgen, R., and Hass, M. 2003., ApJ, 599, L13.
\bibitem[Gillet et al., 1973]{gill73} Gillett, F. C., Forrest, W. J., and Merrill, K. M. 1973, ApJ, 183, 87.
\bibitem[Haas et al., 2003]{haas03} Haas, M, Klaas, U, M\"{u}ller, S. AH, Bertoldi, F, Camenzind, M, Chini, R, Krause, O, Lemke, D, Meisenheimer, K, Richards, P. J, and Wilkes, B. J. 2003. A\&A, 87, 111.
\bibitem[Hatziminaoglou et al., 2008]{hatz08} Hatziminaoglou, E., Fritz, J., Franceschini, A., Afonso-Luis, A., Hern\'an-Caballero, A., P\'erez-Fournon, I., Serjeant, S., Lonsdale, C., Oliver, S., Rowan-Robinson, M., Shupe, D., Smith, H. E., Surace, J. 2008, MNRAS, 386, 1252.
\bibitem[Helou et al., 1985]{helo85}Helou, G, Soifer, B. T, Rowan-Robinson, M. 1985. ApJ, 298, L7.
\bibitem[Horst et al., 2009]{hors09}Horst, H., Duschl, W. J., Gandhi, P., Smette, A. 2009. A\&A, 495, 137
\bibitem[Huchra \&  Burg, 1992]{huch92}Huchra, J., \& Burg, R. 1992. ApJ, 393, 90
\bibitem[Imanishi, 2003]{iman03} Imanishi, M.  2003, ApJ, 599, 918.
\bibitem[Imanishi \& Wada, 2004]{iman04} Imanishi, M and Wada K. 2004, ApJ, 617, 214.
\bibitem[Imanishi, 2006]{iman06} Imanishi, M.  2006, ApJ, 131, 2406.
\bibitem[Imanishi et al., 2006]{iman06b} Imanishi, M., C C Dudley, and Philip R Maloney. 2006, ApJ, 637, 114.
\bibitem[Imanishi et al., 2007]{iman07}Imanishi, M., Dudley, C., Maiolino, R., Maloney, P., Nakagawa, T., \& Risaliti, G.  et. al. 2007, ApJS, 171, 72
\bibitem[Kaspi et al., 2000]{kasp01} Kaspi. S., Smith, P. S., Netzer, H., Maoz, D., Jannuzi, B. T. and Giveon U. 2000, ApJ, 533, 631.
\bibitem[Kaspi et al., 2005]{kasp02} Kaspi. S., Maoz, D., Netzer, H., Peterson, B. M., Vestergaard, M., and Jannuzi, B.T. 2005, ApJ, 629, 61.
\bibitem[Kellermann et al., 1989]{kell05}Kellermann, K., Sramek, R., Schmidt, M., Shaffer, D., \& Green, R. 1989, AJ, 98, 4.
\bibitem[Kennicutt et al., 1998]{kenn98} Kennicutt, Robert C. 1998. Annual Review of Astronomy and Astrophysics, 36, 189.
\bibitem[Khachikian \& Weedman, 1974]{khaw74} Khachikian, E. Ye., \& Weedman, D. W. 1974, ApJ, 192, 581.
\bibitem[Koulouridis et al., 2006]{koul06} Koulouridis, Elias, Chavushyan, Vahram, Plionis, Manolis, Krongold, Yair, Dultzin-Hacyan, Deborah. 2006, ApJ, 651, 93.
\bibitem[Laurent et al., 2000]{laur00}Laurent, O., Mirabel, I. F., Charmandaris, V., Gallais, P., Madden, S. C., Sauvage, M., Vigroux, L., \& Cesarsky, C. 2000, A\&A, 359, 887.
\bibitem[Martini et al., 2001]{mart01}Martini Paul, Pogge, R, Ravindranath, S. \& An, J. H. 2001, 562, 139.
\bibitem[Maiolino et al., 2007]{maio07} Maiolino, R., Shemmer, O., Imanishi, M., Netzer, H., Oliva, E., Lutz, D., and Sturm, E. 2007 A\&A, 468, 979.
\bibitem[Mason et al., 2007]{maso07}Mason, R, Levenson, N. A., Packham, C., Elitzur, M., Radomski, J., Petric, A. O., Wright, G. S. 2007, ApJ, 659, 241.
\bibitem[Mel\'endez et al., 2008]{mele08} Mel\'endez, M, Kraemer, S. B, Schmitt, H. R, Crenshaw, D. M, Deo, R. P, Mushotzky, R. F, Bruhweiler, F. C. 2008, ApJ, 689, 95.
\bibitem[Moorwood et al., 1986]{moor86} Moorwood, A. FM. 1986. A\&A, 166, 4.
\bibitem[Moshir et al., 1990]{mosh90}Moshir, M., et al. 1990, BAAS, 22, 1325.
\bibitem[M\"{u}ller-S\'anchez et al., 2010]{mull10} M\"{u}ller-S\'anchez, F.; Gonz\'alez-Mart\'in, O.; Fern\'andez-Ontiveros, J. A.; Acosta-Pulido, J. A.; Prieto, M. A. 2010. ApJ, 716:1166.
\bibitem[Murayama et al.,  2000]{mura00} Murayama, T., Mouri, H., \& Taniguchi, Y. 2000. ApJ, 528, 179.
\bibitem[Neugebauer et al., 1984]{neug84}Neugebauer, G., et al. 1984, ApJ, 278, L1
\bibitem[O'Neill, 2005]{onei05}O'Neill, P., Nandra, K., Papadakis, and Turner T. 2005. MNRAS, 358,1405.
\bibitem[Osterbrock D., 1978]{oste78} Osterbrock, D. E. 1978. Proc. Natl. Acad. Sci. USA, 75, 540.
\bibitem[Panessa F., 2006]{pane06}Panessa, F., Bassani, L., Cappi, M., Dadina, M., Barcons, X., Carrera, F.J., Ho, L. C. \& Iwasawa, K. 2006, A\&A, 455, 173
\bibitem[Panuzzo et al., 2003]{panu03} Panuzzo, P., Bressan, A., Granato, G. L., Silva, L., and Danese, L. 2003, A\&A, 409, 99.
\bibitem[Peeters et al., 2004]{peet04} Peeters, E., Spoon, H.W.W., and Tielens, A G.G.M.  2004, A\&A, 613, 986.
\bibitem[Peeters, 2002]{peet02}Peeters, E. 2002.  {\it Polycyclic Aromatic Hydrocarbons and dust in regions of massive star formation.} University of Groningen.
\bibitem[Peterson, 1997]{pete97}Peterson B. 1997. {\it An Introduction to Active Galactic Nuclei}. Cambridge University Press.
\bibitem[Pope et al., 2008]{pope08}Pope, Alexandra, Chary, Ranga-Ram, Alexander, David M, Armus, Lee, Dickinson, Mark, Elbaz, David, Frayer, David, Scott, Douglas, Teplitz, Harry. 2008, ApJ, 675, 1171.
\bibitem[Prieto et al., 2005]{prie05}Prieto, M. A., Maciejewski, W., \& Reunanen, J. 2005, AJ, 130, 1472.
\bibitem[Ramos Almeida, 2009]{ramo09} Ramos Almeida C. 2009. {\it Actividad nuclear y formaci\'on estelar en galaxias}. IAC, Universidad de La Laguna.
\bibitem[Read, A. \& Ponman, T. 1998]{read98}Read, A. M., \& Ponman, T. J. 1998, MNRAS, 297, 143.
\bibitem[Rodr\'iguez-Ardila \& Viegas., 2003]{rodr03}Rodr\'{\i}guez-Ardila, A.,\& Viegas, S. M.  2003, MNRAS, 340, L33
\bibitem[Rodr\'iguez-Ardila et al., 2004]{rodr04} Rodr\'iguez-Ardila, A, Pastoriza, M. G, Viegas, S, Sigut, T. A. A, Pradhan, A. K. 2004. A\&A 425, 457.
\bibitem[Rush et al., 1993]{rush93}Rush, B., Malkan, M. A., Spinoglio, L. 1993, ApJSS, 89, 1
\bibitem[Sabater et al., 2008]{saba08} Sabater, J., Leon, S., Verdes-Montenegro, L., Lisenfeld, U., Sulentic, J., \& Verley, S. 2008, A\&A, 486, 73.
\bibitem[Sanders et al., 1988]{sand88} Sanders, D.B., Soifer, B.T., Elias, J.H., Madore, B.F., Matthews, K., Neugebauer, G. and Scoville, N. Z. 1988, ApJ, 325, 74.
\bibitem[Sanders \& Mirabel, 1996]{sand96} Sanders, D. B., Mirabel, I. F. 1996. ARA\&A. 34. 749.
\bibitem[Sanders et al., 2003]{sand03} Sanders, D. B., Mazzarella, J. M., Kim, D.-C., Surace, J. A., \& Soifer, B. T. 2003, AJ, 126, 1607
\bibitem[Seyfert, 1943]{seyf43} Seyfert, C. K. 1943, ApJ, 97, 28.
\bibitem[Shinozaky et al., 2006]{shin06}Shinozaky, K., Miyaji, T., Ishisaki, Y., Ueda, Y., Ogasaka, Y. 2006, ApJ,131, 2843.
\bibitem[Schweitzer et al., 2006]{schw06} Schweitzer, M., Lutz, D., Contursi, A., Tacconi, J., et al. 2006, ApJ, 649, 79.
\bibitem[Soifer et al., 1989]{soif21} Soifer, B. T., Boehmer, L., Neugebauer, G., \& Sanders, D. B. 1989, AJ, 98, 766
\bibitem[Spoon et al., 2001]{spoo01}Spoon, H. W. W., Keane, J. V., Tielens, A. G. G. M., Lutz, D., \& Moorwood,
A. F. M. 2001, A\&A, 365, L353.
\bibitem[Storchi-Bergmann et al., 2005]{stor05} Storchi-Bergmann, T, Nemmen, R. S, Spinelli, P. F, Eracleous, M, Wilson, A. S, Filippenko, A. V, Livio, M. 2005, ApJ, 624, L13.
\bibitem[Spoon et al., 2007]{spoo07} Spoon, H. WW., Marshall, J. A., Houck, J. R., Elitzur, M., Hao, L., Armus, L., Brandl, B. R., and Charmandaris, V. 2007, ApJ, 654, L49.
\bibitem[Storchi-Bergmann, 2008]{stor08} Storchi-Bergmann, Thaisa. 2008, RevMexAA(Serie de Conferencias), 32, 139.
\bibitem[Storchi-Bergmann et al., 2001]{stor01} Storchi-Bergmann, Thaisa., Gonz\'alez Delgado, Rosa M., Schmitt, Henrique R., Cid Fernandes, R. and Heckman, Timothy. 2001, ApJ, 559, 147.
\bibitem[Sturm et al., 2000]{stur00}Sturm, E., Lutz, D., Tran, D., Feuchtgruber, H., Genzel, R., Kunze, D., Moor-
wood, A. F. M., \& Thornley, M. D. 2000, A\&A, 358, 481.
\bibitem[Surace et al., 2001]{sura01} Surace, J., Sanders D. B. \& Evans, A. S. 2001, ApJ 122, 2791.
\bibitem[Taniguchi, 2003]{tani03} Taniguchi, Y. 2003. The Proceedings of the IAU 8th Asian-Pacific Regional Meeting, 289, 353.
\bibitem[Urry \& Padovani, 1995]{urry95}Urry, P. \& Padovani, P. 1995, PASP 107, 803.
\bibitem[Watabe et al., 2008]{wata08} Watabe, Y., Kawakatu, N., and Imanishi, M. 2008, ApJ, 677, 895.
\bibitem[Weedman et al., 2005]{weed05}Weedman, D. W., et al. 2005, ApJ, 633, 706.
\bibitem[Xu et al., 1994]{xu94} Xu C., Lisenfeld U., Volk H.J. 1994, A\&A. 285, 19.
\bibitem[Yong et al., 2007]{yong07} Yong, S., Ogle, P., Rieke, G. H., Antonucci, R., Hines, D. C., et al. 2007, ApJ, 669, 84.
\bibitem[Yun et al., 2001]{yun01} Yun, M. S., Reddy, N. A., \& Condon, J. J. 2001, ApJ, 554, 803.
\end{thebibliography}
\end{document}